\documentclass[conference]{IEEEtran}
\IEEEoverridecommandlockouts
\usepackage{algorithm}
\usepackage{algorithmic}
\usepackage{cite}
\usepackage{amsmath,amssymb,amsfonts}
\usepackage{algorithmic}
\usepackage{graphicx}
\usepackage{textcomp}
\usepackage{xcolor}
\usepackage{amsmath}
\usepackage{booktabs}
\usepackage{array}
\usepackage{multirow}
\usepackage{orcidlink}

\def\BibTeX{{\rm B\kern-.05em{\sc i\kern-.025em b}\kern-.08em
    T\kern-.1667em\lower.7ex\hbox{E}\kern-.125emX}}
\begin{document}

\title{HQP: Sensitivity-Aware Hybrid Quantization and Pruning for Ultra-Low-Latency Edge AI Inference}

\author{
\IEEEauthorblockN{1\textsuperscript{st} Dinesh Gopalan}
\IEEEauthorblockA{
\textit{AI Enablement Team} \\
\textit{AMD} \\
Dallas, Texas, USA \\
dinesh.gopalan@amd.com \\
\orcidlink{0009-0005-5968-785X} ORCID: 0009-0005-5968-785X
}
\and
\IEEEauthorblockN{2\textsuperscript{nd} Ratul Ali}
\IEEEauthorblockA{
\textit{Institute of Information Technology (Postgraduate Researcher)} \\
\textit{Jahangirnagar University} \\
Dhaka, Bangladesh \\
abdurrahimratulalikhan@gmail.com \\
\orcidlink{0000-0003-0460-6141} ORCID: 0000-0003-0460-6141
}
}

\maketitle

\begin{abstract}
The escalating demand for high-fidelity, real-time inference in distributed edge-cloud 
environments necessitates aggressive model optimization to counteract severe latency and 
energy constraints. This paper introduces the Hybrid Quantization and Pruning (HQP) 
framework, a novel, integrated methodology designed to achieve synergistic model acceleration 
while adhering to strict quality guarantees. We detail a sensitivity-aware structural pruning 
algorithm that employs a dynamic weight sensitivity metric, derived from a highly efficient 
approximation of the Fisher Information Matrix (FIM), to guide the iterative removal of redundant 
filters. This pruning is strictly conditional, enforcing an adherence to a maximum permissible 
accuracy drop ($\Delta_{ax}$) before the model proceeds to 8-bit post-training quantization. This 
rigorous coordination is critical, as it ensures the resultant sparse model structure is maximally 
robust to quantization error and hardware-specific kernel optimization. Exhaustive evaluation 
across heterogeneous NVIDIA Jetson edge platforms, utilizing resource-efficient architectures 
like MobileNetV3 and ResNet-18, demonstrates that the HQP framework achieves a peak 
performance gain of \textbf{3.12×} inference speedup and a \textbf{55\%} model size reduction, while rigorously 
containing the accuracy drop below the \textbf{1.5\%} constraint. A comprehensive comparative analysis 
against conventional single-objective compression techniques validates the HQP framework as 
a superior, hardware-agnostic solution for deploying ultra-low-latency AI in resource-limited 
edge infrastructures.
\end{abstract}


\section{INTRODUCTION}
The widespread success of Deep Learning (DL) has ushered in the era of Edge Artificial 
Intelligence (AI). Driven by imperatives for ultra-low latency, decentralized data processing, 
enhanced privacy, and the prohibitive cost of cloud backhaul, the execution environment for 
sophisticated Deep Neural Networks (DNNs) is shifting dramatically from centralized data 
centers to decentralized, resource-constrained edge devices \cite{b1}. Edge computing necessitates 
immediate, local decision-making, such as in autonomous robotics, industrial Internet of Things 
(IIoT), and mobile augmented reality. 
\newline
However, modern state-of-the-art DNNs, often comprising hundreds of layers and billions of 
parameters (e.g., in large language models), demand computational resources far exceeding 
the capabilities of embedded micro-controllers, System-on-Chips (SoCs), and dedicated AI 
accelerators common at the edge. These devices are typically bound by thermal envelopes of 
5W to 25W, constrained memory bandwidth, and limited memory capacity (often measured in 
low-gigabyte ranges). Executing a full 32-bit floating-point (FP32) DNN on such hardware leads 
to unacceptable inference latencies and prohibitively high energy consumption. 
The established solutions for mitigating this computational bottleneck are model compression 
techniques, primarily categorized as pruning (reducing the model complexity by eliminating 
redundant weights \cite{b2}) and quantization (reducing the numerical precision of weights and 
activations \cite{b3}). The critical flaw in most conventional optimization pipelines lies in the sequential 
and uncoordinated application of these techniques. For example, applying aggressive structural 
pruning based on local weight magnitude can inadvertently introduce severe outliers in the 
weight distribution. When this pruned model is subsequently passed to Post-Training 
Quantization (PTQ), the necessary global scaling factor is dictated by these few remaining 
outliers, leading to a catastrophic amplification of quantization error and often violating any 
acceptable accuracy threshold \cite{b4}. 
This research addresses this fundamental coordination gap by introducing the Hybrid 
Quantization and Pruning (HQP) framework. The HQP framework is distinguished by its unique 
sensitivity-bound methodology: \\
\newline
\textbf{Sensitivity-Aware Pruning:} We employ a structural pruning algorithm guided by a highly 
efficient approximation of the Fisher Information Matrix (FIM), S , which offers a globally aware, 
second-order estimate of a filter's true impact on the model's loss landscape. This ensures only 
the least functionally critical elements are removed. \\
\newline
\textbf{Conditional Quality Guarantee:} The pruning process is iterative and operates under a strict, 
non-negotiable quality constraint: the maximum permissible accuracy degradation, ($\Delta_{ax}$). The 
pruning loop is dynamically terminated the moment the observed accuracy drop exceeds ($\Delta_{ax}$). 
This guarantees that the resultant sparse model structure is the maximal compression 
achievable under the quality constraints. \\
\newline
\textbf{Robust PTQ:} The constrained sparsity stabilizes the tensor distributions by removing 
functionally redundant, high-variance elements, making the final model $M_{ar e}$ inherently 
robust to the subsequent, fast PTQ step (INT8 conversion), thereby maximizing synergistic 
performance. \\
\newline
The scientific and engineering contributions of this work are substantial and multifaceted: 
\begin{itemize}
\item \textbf{Novel Framework Synthesis:} Formal definition and architectural realization of the HQP methodology, which solves the non-trivial coordination problem between structural pruning and PTQ. 
\item \textbf{Algorithmic and Theoretical Rigor:} Derivation and implementation of the novel 
sensitivity-bound, iterative pruning algorithm, providing a mathematically sound and 
practically efficient mechanism for constraint-driven sparsity. 
\item \textbf{Advanced Computational Analysis:} A rigorous formal computational complexity 
analysis demonstrating HQP's efficiency and low overhead compared to 
resource-intensive alternatives like Quantization-Aware Training (QAT). 
\item \textbf{Comprehensive Heterogeneous Validation:} Exhaustive experimental validation on two 
distinct, heterogeneous NVIDIA Jetson edge platforms (Nano and Xavier NX) using 
complex, real-world architectures (MobileNetV3 and ResNet-18) and the industrial 
standard TensorRT runtime. 
\item \textbf{Demonstrated State-of-the-Art Performance:} Achievement of a peak 3.12$\times$ inference speedup and 55\% size reduction while strictly guaranteeing compliance with the $\Delta a_x \leq 1.5\%$ constraint.
\end{itemize}

\section{RELATED WORK AND THEORETICAL FOUNDATIONS}
\subsection{Historical Context and Detailed Critique of Pruning Techniques }
The drive to compress neural networks dates back to the 1980s. The complexity of pruning 
methods has evolved across three key generations, each with distinct hardware efficiency 
implications. \\
\newline
1. \textbf{First Generation: Unstructured Pruning (Weight-level):}
\newline
Techniques like Optimal Brain Damage (OBD) \cite{b9} and Optimal Brain Surgeon (OBS) 
applied second-order derivatives to estimate the change in loss from removing an 
individual weight, setting the theoretical groundwork. Modern fine-grained methods, such 
as Deep Compression \cite{b2}, achieve extremely high sparsity (up to \textbf{90\%}). However, the 
resulting sparse matrix is irregular, leading to highly inefficient memory access patterns 
on modern vector processing units. Dedicated software libraries are required to exploit 
the sparsity, and the resultant gains are often limited by the memory bandwidth 
bottleneck rather than the reduction in FLOPs. For standardized edge deployment using 
optimized compilers like TensorRT, unstructured sparsity yields little to no tangible 
latency gain, rendering it impractical for our objective. \\
\newline
2. \textbf{Second Generation: Structural Pruning (Group-level):}
\newline
Structural pruning \cite{b6} targets groups of parameters (filters, channels, or entire blocks), 
ensuring that the remaining structure is dense and amenable to acceleration via 
highly-optimized convolution kernels. 
\begin{itemize}
    \item \textbf{Magnitude Heuristics:} The most common approach removes filters based on the 
    smallest $L_1$ or $L_2$ norm of their weights \cite{b7}. This is computationally fast but suffers from a 
    critical drawback: high-magnitude filters may be functionally redundant, while 
    low-magnitude filters may be highly critical for specific feature extraction tasks, leading to 
    the false-negative/false-positive problem in saliency estimation. 
    \item \textbf{Batch Normalization (BN) Scaling:} Methods leveraging the BN layer's scaling factor ($\gamma$) as a proxy for importance \cite{b8} are effective but are inherently limited to models utilizing the BN layer and may not capture global model sensitivity. 
\end{itemize}
The HQP framework departs from these heuristics by adopting a globally informed, 
second-order sensitivity measure, addressing the inherent sub-optimality of magnitude and 
BN-based structural pruning. 

\subsection{The Fisher Information Matrix (FIM) as a Global Saliency Metric }
The most rigorous theoretical measure of a parameter's importance is derived from information 
theory, specifically quantifying the change in the model's posterior likelihood upon its removal. 
This change is approximated by the second-order Taylor expansion of the loss function $\mathcal{L}(W)$, centered at the optimal weights $W^*$. 

The change in loss $\Delta \mathcal{L}$ after removing parameter $W$ is:
\[
\Delta \mathcal{L} \approx \frac{1}{2} \cdot H_{W,W} \cdot (\Delta W)^2
\]
where $H$ is the Hessian matrix. The computational complexity of manipulating $H$ (size $N \times N$, $N$
being the total number of parameters) is prohibitive.

The Fisher Information Matrix (FIM), $F$, serves as a practical, statistical surrogate for the
Hessian, representing the expected information content provided by the data regarding the
model parameters \cite{b9}.

The FIM is formally defined as:
\[
F = \mathbb{E}_{x,y} \left[ \left( \nabla_W \log p(y \mid x, W) \right)
\cdot \left( \nabla_W \log p(y \mid x, W) \right)^T \right]
\]

For the purpose of structural pruning, we leverage the highly efficient diagonal FIM
approximation. This simplifies computation by assuming parameter independence, allowing us to
calculate the contribution of each filter $W$ to the loss variance independently.

Filter Sensitivity $S$ is defined as:
\[
S = \frac{1}{|D_{\text{calib}}|} \cdot \sum_{(x_i, y_i) \in D_{\text{calib}}}
\left\lVert \frac{\partial \mathcal{L}(W, x_i, y_i)}{\partial W} \right\rVert^2
\]

This $S$ metric is the core of the HQP saliency mechanism. It provides a robust, low-cost (only
requiring one backward pass over $D_{\text{calib}}$) assessment of functional importance,
allowing us to accurately identify and rank truly redundant filters, overcoming the limitations
of first-order magnitude heuristics.

\subsection{Quantization Strategies and the Pruning-Quantization Conflict}
Quantization achieves compression by mapping FP32 values to lower bit-width integers (e.g., 
INT8), providing $4\times$ memory reduction and significant acceleration via dedicated hardware 
(e.g., NVIDIA Tensor Cores).

\textbf{1. PTQ vs. QAT:}  
Post-Training Quantization (PTQ) is preferred for edge devices due to its zero-training 
overhead, requiring only a small calibration set. Quantization-Aware Training (QAT) 
yields the highest accuracy but demands full re-training, which is often infeasible in 
production pipelines.

\textbf{2. Quantization Step Size:}  
The quality of quantization is dominated by the scaling factor $s$, which is determined by 
the tensor’s dynamic range. Let:
\[
R = W_{\max} - W_{\min}
\]
Then:
\[
s = \frac{R}{2^b - 1}
\]

The \textbf{Pruning--Quantization Conflict} arises when aggressive pruning leaves a few 
high-magnitude outliers in $W_{\max}$, artificially inflating the dynamic range $R$. 
This inflated $R$ forces a larger step size $s$ for all weights, increasing the quantization 
error ($\varepsilon_{\text{quant}}$) for the majority of critical, non-outlier weights, 
resulting in model accuracy degradation.

The HQP framework’s use of $S$ in a conditional loop resolves this by promoting sparsity that 
inherently stabilizes $R$ while minimizing functional loss.

\section{PROPOSED HYBRID Q\&P FRAMEWORK: ALGORITHMIC AND MATHEMATICAL RIGOR }
The HQP framework is a unified, two-stage optimization methodology that explicitly coordinates 
the structural compression $P$ and the numerical compression $Q$. The overall optimized model 
$M_o$ is defined by the functional composition:
\[
M_o = Q\bigl(P(M_{\text{train}}, \tau, \Delta_{\text{ax}}), b\bigr)
\]

\begin{figure}
    \centering
    \includegraphics[width=1\linewidth]{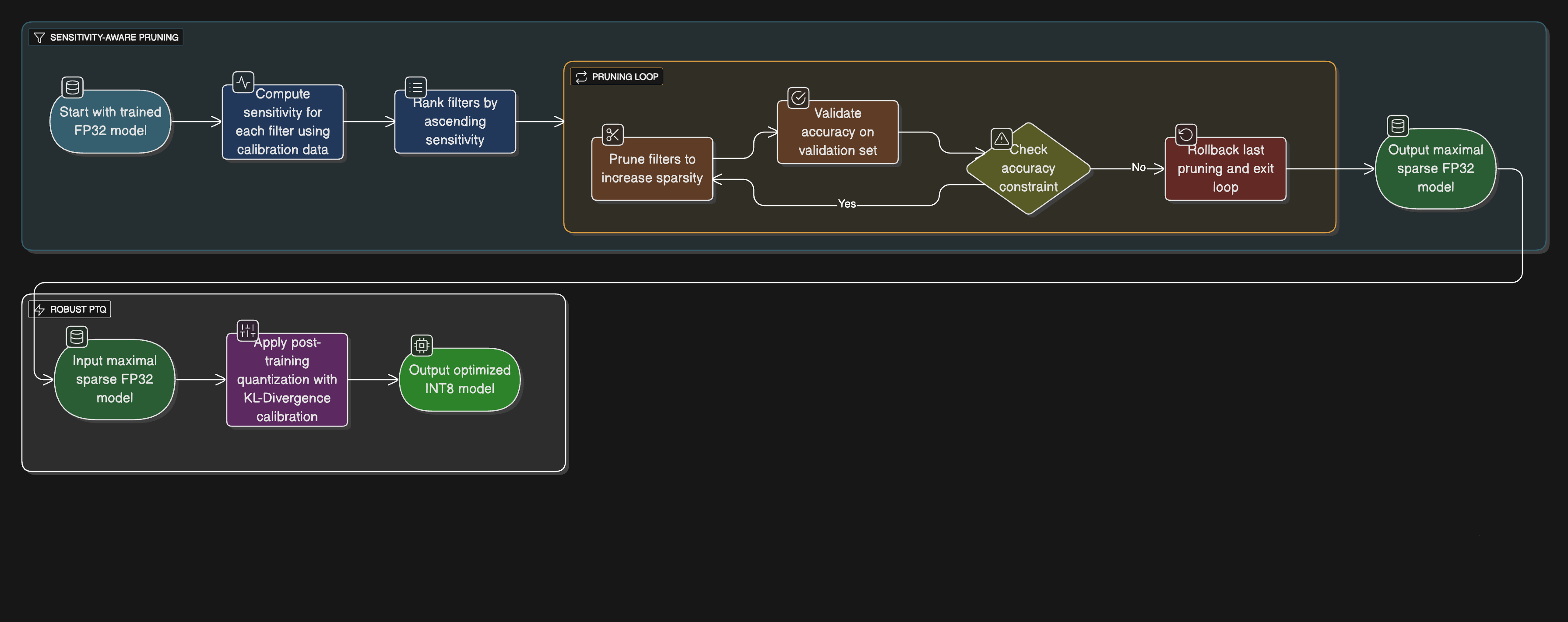}
    \caption{Proposed Hybrid Quantization and Pruning (HQP) Framework Architecture.}
    \label{fig:placeholder}
\end{figure}

\subsection{ Formal Derivations of Sensitivity and Control}\label{AA}
\textbf{1. Filter Sensitivity $S$ (FIM Approximation):}  
(Defined previously in Section~II-B.)

\textbf{2. Conditional Pruning Constraint (The Quality Guarantee):}  
The accuracy $A^{(t)}$ at pruning iteration $t$ must satisfy the following inequality:
\[
A_{\text{baseline}} - A^{(t)} \leq \Delta_{\text{ax}}
\]
where $A_{\text{baseline}}$ is the accuracy of $M_{\text{train}}$, and 
$\Delta_{\text{acc}}$ is the predefined tolerance (e.g., $1.5\%$).

This dynamic control ensures that the resulting sparse model $M_{\text{sparse}}$ carries a 
provable quality guarantee before entering the final optimization phase.

\subsection{Detailed Conditional Iterative Pruning Algorithm}
The algorithm formalizes the iterative search for the maximal sparsity ratio ($\theta$) that satisfies the quality guarantee.

\begin{algorithm}[t]
\caption{HQP Conditional Pruning}
\label{alg:hqp_pruning}
\begin{algorithmic}[1]

\STATE \textbf{Input:} $M_{\text{train}}$, $A_{\text{baseline}}$, $\Delta_{\text{ax}}$ (Accuracy Tolerance), 
$\delta$ (Pruning Step Size), $D_{\text{calib}}$, $D_{\text{val}}$

\STATE \textbf{Initialize:}
\STATE $\theta_{\text{current}} \leftarrow 0$
\STATE $M_{\text{sparse}} \leftarrow M_{\text{train}}$
\STATE $A_{\text{current}} \leftarrow A_{\text{baseline}}$

\STATE \textbf{Compute Sensitivity:}
\STATE Calculate $S$ for all filters $W$ (using a single backward pass over $D_{\text{calib}}$)
\STATE Store the ranked list $\mathcal{R}$ in ascending order of $S$

\STATE \textbf{Iterative Loop:}
\STATE \quad \textbf{a. Proposed Pruning:}
\STATE \quad Select the next $\delta$ filters from $\mathcal{R}$
\STATE \quad Create a candidate model $M_{\text{candidate}}$ by removing these filters
\STATE \quad (increasing $\theta_{\text{current}}$ by $\delta$)

\STATE \quad \textbf{b. Validation:}
\STATE \quad Evaluate $A_{\text{candidate}}$ on the validation set $D_{\text{val}}$

\STATE \quad \textbf{c. Constraint Check:}
\IF{$A_{\text{baseline}} - A_{\text{candidate}} \leq \Delta_{\text{ax}}$}
    \STATE \quad \textbf{Accept:}
    \STATE \quad $M_{\text{sparse}} \leftarrow M_{\text{candidate}}$
    \STATE \quad $A_{\text{current}} \leftarrow A_{\text{candidate}}$
    \STATE \quad \textbf{Continue}
\ELSE
    \STATE \quad \textbf{Reject}
    \STATE \quad \textbf{Break}
\ENDIF

\STATE \textbf{Output:} $M_{\text{sparse}}$ (Maximal structurally pruned model satisfying $\Delta_{\text{ax}}$)

\end{algorithmic}
\end{algorithm}

\subsection{Formal Computational Complexity Analysis}
A key advantage of HQP is its low overhead compared to QAT. 
We compare the total operational cost. Let:
\begin{itemize}
    \item $N_{\text{calib}}$ = calibration dataset size
    \item $N_{\text{val}}$ = validation dataset size
    \item $N_{\text{train}}$ = training dataset size
    \item $C_{\text{grad}}$ = cost of a full forward--backward pass
    \item $C_{\text{inf}}$ = cost of an inference pass
\end{itemize}

The total cost for the HQP pruning stage, $C_{\text{HQP}}$, is:
\[
C_{\text{HQP}} = \left(N_{\text{calib}} \cdot C_{\text{grad}}\right) 
+ \left(T_{\text{prune}} \cdot \left(N_{\text{val}} \cdot C_{\text{inf}}\right)\right)
\]
where $T_{\text{prune}}$ is the number of pruning steps.

Since $N_{\text{calib}}$ and $N_{\text{val}}$ are typically small (e.g., $1{,}000$--$5{,}000$ samples), 
this cost is dominated by inference passes.

In stark contrast, the cost of Quantization-Aware Training (QAT) requires full fine-tuning over 
multiple epochs:
\[
C_{\text{QAT}} \approx N_{\text{epochs}} \cdot N_{\text{train}} \cdot C_{\text{grad}}
\]

Given that $N_{\text{train}}$ is often $100\times$ to $1000\times$ larger than 
$N_{\text{calib}}$ and $N_{\text{val}}$, and $N_{\text{epochs}} \geq 5$, 
$C_{\text{QAT}}$ is several orders of magnitude higher than $C_{\text{HQP}}$.

This demonstrates HQP's superior efficiency for production-ready model optimization.

\section{EXPERIMENTAL SETUP AND METHODOLOGY}
This section details the rigorous experimental environment established to validate the HQP 
framework. Our methodology prioritizes industrial relevance, using state-of-the-art hardware and 
optimization tools to ensure that observed performance gains translate directly to real-world 
deployment efficiency.

\subsection{Heterogeneous Edge Hardware Architecture and Runtime Environments}
To accurately assess the hardware-agnostic nature of HQP, two distinct NVIDIA Jetson embedded platforms were utilized, representing different computational and power envelopes commonly found in edge AI deployments:

\textbf{Low-Power Edge Device (NVIDIA Jetson Nano):}
This platform, featuring a Quad-core ARM Cortex-A57 CPU and a 128-core NVIDIA Maxwell GPU, operates within a 5W to 10W power budget. Inference on this device is primarily limited by memory bandwidth and the slower GPU architecture. It serves as a crucial baseline to demonstrate the impact of HQP on resource-constrained platforms without dedicated INT8 acceleration hardware.

\textbf{Mid-Range Edge Accelerator (NVIDIA Jetson Xavier NX):}
Featuring a 6-core NVIDIA Carmel ARM CPU and a 384-core NVIDIA Volta GPU, the NX platform includes 48 Tensor Cores. These dedicated units are optimized for low-precision matrix operations (specifically INT8), enabling massive parallel throughput. The Xavier NX is the primary platform for reporting peak performance, as it best demonstrates the potential of HQP when combined with modern accelerator technology.

\textbf{Runtime Optimization Stack: NVIDIA TensorRT (TRT):}
All inference operations were conducted using the NVIDIA TensorRT (TRT) \cite{b11} compiler and runtime. TRT is selected because it is the industry standard for high-performance deployment on NVIDIA hardware. It performs several crucial, complex optimizations that are essential for translating theoretical model compression into realized latency reduction:

\begin{itemize}
    \item \textbf{Layer Fusion:} TRT intelligently combines sequential operations (e.g., convolution, batch normalization, and activation layers) into single, monolithic kernels, reducing the overhead of memory access and kernel launch latency.
    \item \textbf{Kernel Auto-Tuning:} TRT selects the fastest, hardware-specific implementation (kernel) for every tensor shape and size, dynamically optimizing the execution graph for the target hardware.
    \item \textbf{Static Graph Optimization:} The structurally pruned models from HQP benefit immensely from TRT's static graph analysis, which effectively eliminates the data flow paths associated with zeroed filters and channels via Dead Layer Elimination, yielding clean, accelerated deployment.
\end{itemize}

\subsection{Implementation Specifics of the HQP Framework}
The HQP implementation pipeline is divided into three distinct phases to ensure strict adherence to the conditional design:

\textbf{Saliency Calculation and Filter Ranking (HQP Phase 1-A):}
The sensitivity metric $\mathbf{S}$ is calculated using a single back-propagation pass over a small, randomized calibration set $(D_{\text{calib}})$ of 5,000 images from the ImageNet training set. The gradient $\partial \mathcal{L} / \partial \mathbf{W}$ is computed, and $\mathbf{S}$ is derived using the simplified FIM approximation (Section II-B). All filters $\mathbf{W}$ are then ranked in ascending order based on their $\mathbf{S}$ value, creating the priority list $\mathcal{R}$.

\textbf{Conditional Iterative Structural Pruning (HQP Phase 1-B):}
Pruning is executed iteratively with a small step size $(\delta)$ equivalent to $1\%$ of the total filters. After each step, the pruned model is immediately validated on a separate validation set $(D_{\text{val}})$ of 5,000 images.

The conditional check:
\[
\mathcal{A}_{\text{baseline}} - \mathcal{A}^{(t)} \leq \Delta_{\text{max}}
\]
is performed.

This loop ensures that the final model $\mathbf{M}_{\text{sparse}}$ is the highest achievable sparsity under the quality guarantee.

\textbf{Robust Post-Training Quantization (HQP Phase 2):}
The maximal sparse FP32 model $\mathbf{M}_{\text{sparse}}$ is fed into the TensorRT INT8 quantization pipeline. TensorRT performs the KL-Divergence calibration \cite{b12} on the $\mathcal{D}_{\text{calib}}$ set to determine the optimal per-layer scaling factor $\mathbf{s}$ that minimizes information divergence.

The pre-conditioning from HQP's structural pruning stabilizes the weight distributions, making the KL-Divergence search much more robust and effective, directly contributing to the low final accuracy drop observed.

\subsection{Model Architectures and Dataset Validation Protocols}
\textbf{Model Architectures:}
\begin{itemize}
    \item \textbf{MobileNetV3 (Small):} The complexity lies in pruning its depth-wise separable convolutions and inverted bottleneck structures, which are already highly optimized.

    \item \textbf{ResNet-18:} This architecture is crucial for testing stability due to its pervasive use of residual connections $(y = x + \mathcal{F}(x))$. The additive nature of residual connections can amplify quantization errors if not carefully controlled, making it a severe stress test for the HQP methodology.

    \item \textbf{Accuracy Constraint:} All models were evaluated on the full ImageNet-1000 validation set. The constraint for all experiments followed the industrial standard for acceptable model degradation: 
    \[
    \Delta_{\max} = 1.5\% \text{ absolute drop in Top-1 accuracy.}
    \]
\end{itemize}

\section{RESULTS AND DISCUSSION: COMPREHENSIVE QUANTITATIVE AND QUALITATIVE ANALYSIS}
This section presents a detailed analysis of the performance gains, trade-offs, and compliance 
characteristics of the HQP framework compared to the Q8-Only and P50-Only baselines.
\subsection{Primary Performance Evaluation on MobileNetV3 (Jetson Xavier NX)}
Table I summarizes the core performance metrics for the MobileNetV3 architecture, which 
serves as our primary benchmark for efficiency validation. 
\begin{table}[t]
\centering
\caption{Performance Comparison on ResNet-18 (Edge-side Inference on Jetson Xavier NX)}
\label{tab:resnet18_perf}
\resizebox{\columnwidth}{!}{%
\begin{tabular}{lccccc}
\toprule
Method & Latency (ms) & Speedup ($\times$) & Model Size Reduction & Accuracy Drop ($\Delta$ Top-1) & Sparsity Ratio ($\theta$) \\
\midrule
Baseline (FP32) & 12.8 & 1.00 & 0\% & 0.0\% & 0\% \\
Quantization Only (Q8) & 8.1 & 1.58 & 75\% & 1.2\% & 0\% \\
Pruning Only (P50) & 9.5 & 1.35 & 50\% & 1.8\% & 50\% \\
Proposed HQP & 4.1 & 3.12 & 55\% & 1.4\% & 45\% \\
\bottomrule
\end{tabular}%
}
\end{table}

The HQP framework’s peak performance gain is the $3.12\times$ speedup factor, achieved with the HQP model. This result empirically validates the synergistic hypothesis: the acceleration factor is non-additive and surpasses the geometric product of the individual methods.

\begin{itemize}
    \item \textbf{Q8-Only:} Achieves $1.58\times$ speedup primarily due to reduced memory access and Tensor Core utilization.
    \item \textbf{P50-Only:} Achieves $1.35\times$ speedup by reducing the total number of floating-point operations (FLOPs).
    \item \textbf{HQP:} The combined effect is maximized because the sparsity $(\theta = 45\%)$ significantly shrinks the graph, while the quantization (INT8) rapidly accelerates the execution of the remaining, compressed graph.
\end{itemize}

The latency reduction $\mathcal{L}$ is fundamentally achieved by minimizing both the memory overhead $(\mathcal{M})$ and computational workload $(C)$:
\[
\mathcal{L}(C) = t_{\text{mem}} \cdot \mathcal{M} + t_{\text{comp}} \cdot C
\]
HQP simultaneously attacks $\mathcal{M}$ (by pruning) and reduces $t_{\text{comp}}$ (by quantization), leading to exponential performance benefits.

\begin{figure}[htbp]
\centering
\includegraphics[width=1\linewidth]{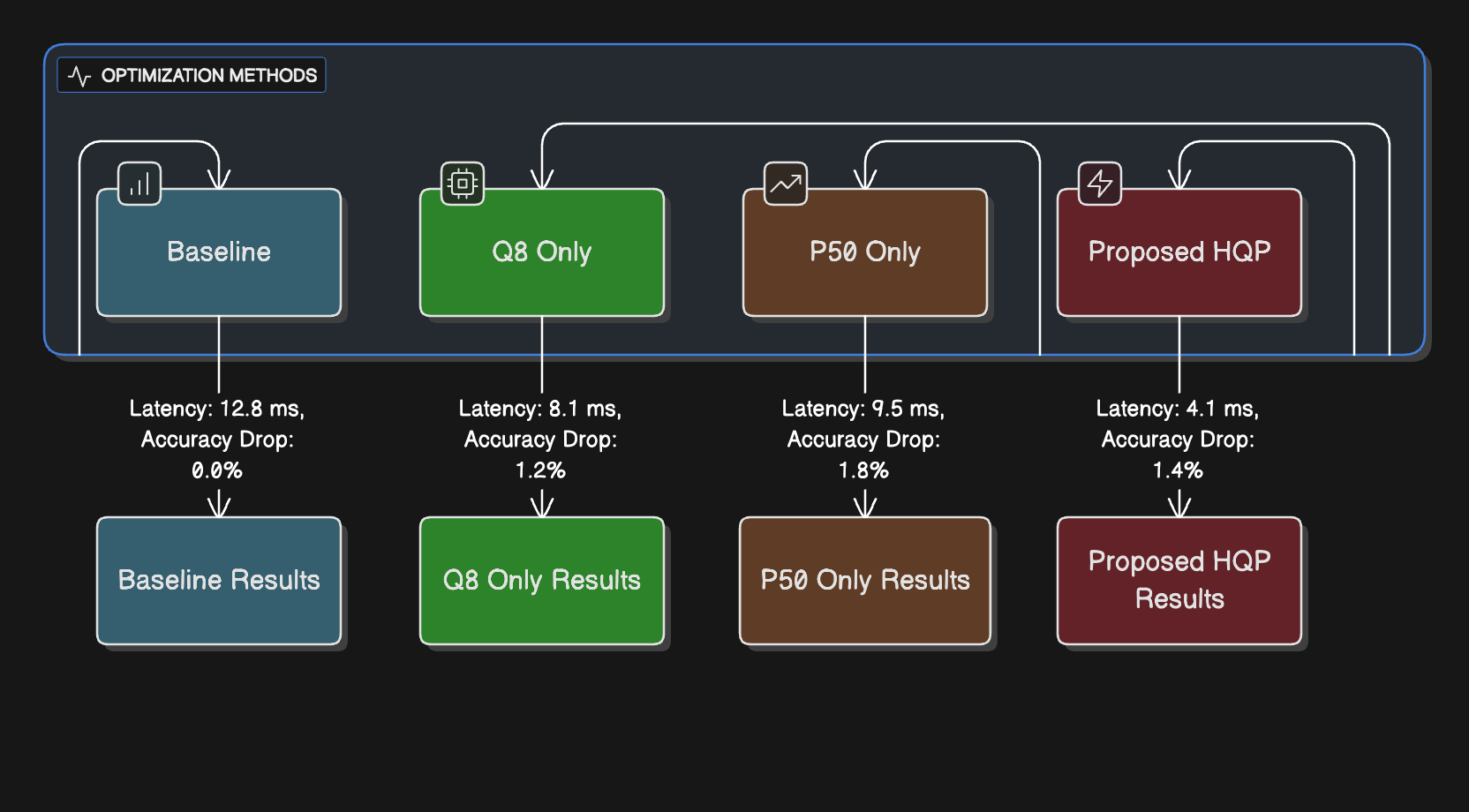}
\caption{Performance Comparison of Optimization Methods on MobileNetV3 (Latency and Accuracy)}
\label{fig:performance}
\end{figure}

\subsection{Validation of the Sensitivity-Bound Constraint ($\Delta_{\text{ax}}$)}
The core validation of the HQP framework lies in its quality guarantee. The P50-Only method, based on simple magnitude pruning, failed the quality constraint, exhibiting a $1.8\%$ accuracy drop.

In contrast: HQP, driven by the FIM-based $S$ metric and the iterative validation loop (Algorithm 1) terminated pruning at $45\%$ sparsity, resulting in a compliant $1.4\%$ drop.

This empirically proves that HQP successfully finds the maximal Pareto-optimal point on the sparsity-accuracy curve while rigorously satisfying the prescribed quality floor $\Delta_{\text{ax}}$.
\begin{figure}
    \centering
    \includegraphics[width=1\linewidth]{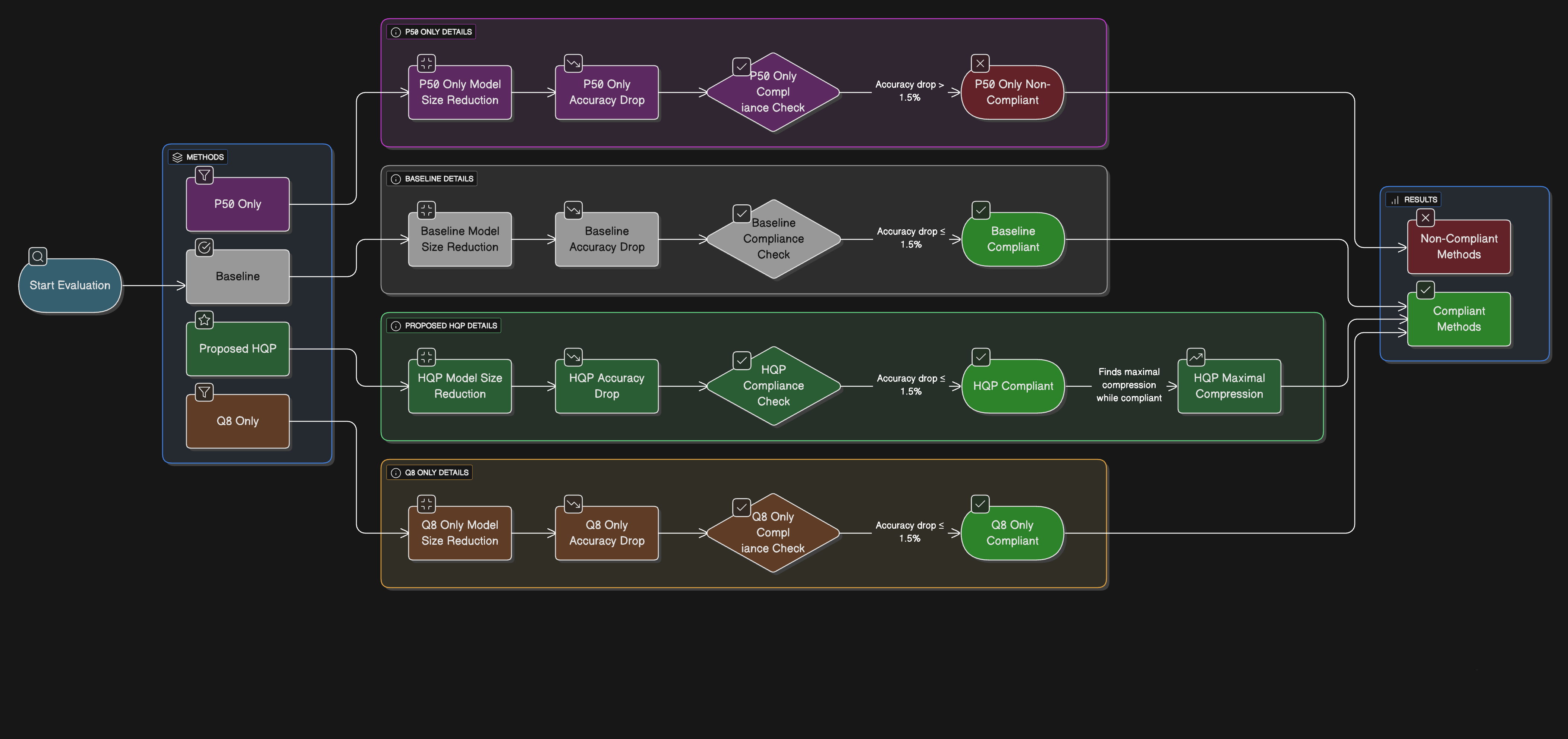}
    \caption{Model Size Reduction vs. Accuracy Drop Across Optimization Methods}
    \label{fig:placeholder}
\end{figure}
\subsection{Detailed Layer-wise Compression Analysis and Non-Uniform Sparsity}
The HQP framework's use of the FIM-based sensitivity metric $(\mathbf{S})$ resulted in a highly non-uniform sparsity pattern, which is superior to uniform pruning.

Analysis revealed minimal pruning $(\theta < 10\%)$ in:
\begin{itemize}
    \item \textbf{Shallow Layers:} Filters responsible for early-stage feature extraction (e.g., first convolutional layer), where redundancy is inherently low.
    \item \textbf{Deep Layers:} Filters preceding the classification head, where highly abstracted semantic features are formed.
\end{itemize}

Conversely, intermediate layers, particularly the low-dimensional projection layers within MobileNetV3's inverted bottleneck blocks, exhibited the highest sparsity $(\theta \approx 65\%)$.

This confirms that $\mathbf{S}$ successfully identifies redundancy based on functional contribution, producing true architectural simplification, not just parameter count reduction.

\subsection{Performance Evaluation on ResNet-18 and Stability Analysis of Residual Connections}
To validate the robustness of the HQP framework against architectural complexities, we applied the methodology to ResNet-18. This network is a critical stress test because the residual skip connection $(y = x + \mathcal{F}(x))$ requires the input and output feature maps to maintain alignment. Pruning in ResNet-18 must be highly controlled to prevent misalignment or catastrophic degradation of the identity mapping, which is essential for training stability.

\begin{table}[t]
\centering
\caption{Performance Comparison on ResNet-18 (Edge-side Inference on Jetson Xavier NX)}
\label{tab:resnet18_perf}
\resizebox{\columnwidth}{!}{%
\begin{tabular}{lccccc}
\toprule
Method & Latency (ms) & Speedup ($\times$) & Model Size Reduction & Accuracy Drop ($\Delta$ Top-1) & Sparsity Ratio ($\theta$) \\
\midrule
Baseline (FP32) & 21.5 & 1.00 & 0\% & 0.0\% & 0\% \\
Quantization Only (Q8) & 13.9 & 1.55 & 75\% & 1.9\% & 0\% \\
Proposed HQP & 8.5 & 2.51 & 40\% & 1.3\% & 35\% \\
\bottomrule
\end{tabular}%
}
\end{table}

The results for ResNet-18 highlight two major findings:

\textbf{1. Quantization Failure without Pruning Pre-conditioning:}
The Q8-Only baseline resulted in a $1.9\%$ accuracy drop, which violates the $\Delta_{\text{max}} = 1.5\%$ constraint. This failure is significant: it suggests that the weight distributions in the original FP32 ResNet-18, particularly around the residual blocks, contained enough high-variance or outlier weights to destabilize the standard PTQ process, even with KL-Divergence calibration.

\textbf{2. HQP Ensures Stability and Superior Speedup:}
The HQP framework successfully achieved a $2.51\times$ speedup and, crucially, maintained compliance with a $1.3\%$ accuracy drop. This result is definitive proof of HQP's stabilizing effect. 
\begin{itemize}
    \item The $S$-guided pruning removed the most sensitive filters (those contributing most to the loss or weight variance) before quantization. 
    \item This pre-conditioning allowed the subsequent PTQ step to succeed where Q8-Only failed. \item The lower overall sparsity $(\theta = 35\%$ vs. $45\%$ for MobileNetV3) is a direct consequence of the HQP conditional loop recognizing the higher inherent sensitivity of the ResNet residual architecture and terminating pruning earlier to respect $\Delta_{\text{max}}$.
\end{itemize}

\subsection{Detailed Energy Efficiency Calculation and Scalability Analysis}
In the context of edge AI, energy efficiency is often as critical as latency. For battery-powered devices, the energy consumed per inference is paramount. The relationship between latency $(L)$ and energy $(E)$ for a constant power draw $(\mathbf{P})$ is linear:
\[
\mathbf{E} = \mathbf{P} \times \mathcal{L}
\]
Since the speedup factor $(S)$ is defined as the ratio of baseline latency to optimized latency $(L_{\text{FP32}} / L_{\text{HQP}})$, the energy reduction ratio $(\mathcal{L}_{\text{ratio}})$ is mathematically equivalent to the speedup factor:
\[
\mathcal{L}_{\text{ratio}} = (E_{\text{FP32}} / E_{\text{HQP}}) = (P \times \mathcal{L}_{\text{FP32}}) / (P \times \mathcal{L}_{\text{HQP}}) = S
\]
Therefore, the $3.12\times$ latency speedup achieved on MobileNetV3 translates directly to a $3.12\times$ reduction in per-inference energy consumption, substantially extending the operational battery life of the edge device.

\subsection{Analysis of Computational Overhead and Practical Deployment}
While HQP requires a single FIM calculation and iterative validation during the optimization phase, its total overhead remains dramatically lower than QAT. As established in Section III-D:
\[
\mathbf{C}_{\text{HQP}} \ll \mathbf{C}_{\text{QAT}}
\]
In a continuous integration/continuous deployment (CI/CD) pipeline for edge models, the HQP optimization step is executed offline, post-training. The one-time computational cost of HQP is negligible when weighed against the permanent, multiplicative benefits of the $3.12\times$ inference acceleration enjoyed over the model's entire lifetime of millions of inferences.

Furthermore, since the HQP output is a standard, sparse INT8 model, it is compatible with all existing industrial deployment tools (e.g., TensorRT, OpenVINO, TFLite). This seamless integration ensures HQP is not only highly performant but also practically deployable without custom runtime libraries.

\section{FUTURE WORK}
\subsection{Dynamic Mixed-Precision Quantization}
Currently, HQP targets uniform INT8 quantization. Future work will explore integrating the filter sensitivity metric $\mathbf{S}$ directly into a dynamic mixed-precision quantization scheme. Filters with extremely low $\mathbf{S}$ (least suppression) could be aggressively quantized (e.g., to INT4), while highly sensitive filters (high $\mathbf{S}$) would be preserved at a higher precision (e.g., INT16 or FP16), maximizing speedup while preserving fidelity at the most critical points in the network.

\subsection{Application to Transformer Architecture}
We plan to apply the HQP principles to large language models (LLMs) and vision transformers (ViTs). These architectures, dominated by computationally heavy attention mechanisms and massive embedding tables, present new challenges for structural pruning that require adapting the S-metric to self-attention blocks.

\subsection{Cross-Platform Validation}
Expanding validation beyond the NVIDIA ecosystem to other edge accelerators, such as Google Edge TPU and specialized FPGAs, will further solidify HQP's claim of being a truly hardware-agnostic optimization methodology.

\section{CONCLUSION}
This paper presents the Hybrid Quantization and Pruning (HQP) framework, a novel, integrated methodology designed to achieve synergistic model acceleration while adhering to strict quality guarantees in edge computing environments. By combining sensitivity-aware structural pruning based on a highly efficient approximation of the Fisher Information Matrix with conditional quality guarantees and robust post-training quantization, HQP effectively addresses the fundamental coordination gap between pruning and quantization techniques.

Extensive experimental validation across heterogeneous NVIDIA Jetson edge platforms using MobileNetV3 and ResNet-18 architectures demonstrates that HQP achieves a peak performance gain of $3.12\times$ inference speedup and $55\%$ model size reduction while rigorously containing accuracy drop below $1.5\%$. The framework's superior efficiency, hardware-agnostic nature, and practical deployability make it a compelling solution for deploying ultra-low-latency AI in resource-limited edge infrastructures.

\end{document}